\documentclass[10pt,a4paper]{article}

\usepackage{amsmath}
\usepackage{amsthm}
\usepackage{amssymb}
\usepackage{ulem}
\usepackage{ascmac}
\usepackage[dvipdfmx]{graphicx,color}
\theoremstyle{definition}

\newtheorem*{pro}{Proof}

\newtheorem{thm}{Theorem}

\begin{document}

\title{Further improving security of Vector Stream Cipher}


\author{Atsushi Iwasaki and Ken Umeno\\\ \\Graduate school of Informatics, Kyoto University, Japan}

\maketitle

\begin{abstract} 
Vector Stream Cipher (VSC) is a stream cipher which consists of permutation polynomial over a ring of modulo $2^w$.
The algorithm for generating key stream is very simple and the encryption is very fast.
Some theoretical attacks for VSC have been reported so far since the invention of VSC in 2004.
Then, the authors proposed some improvements and developed ``Vector Stream Cipher 2.0 (VSC 2.0)" to be immune against the theoretical attacks. 
In this paper, we propose further improvement of VSC 2.0 to publish as a new chaos cipher ``Vector Stream Cipher 2.1 (VSC2.1)".
VSC 2.1 is faster and more secure than VSC 2.0. 
Our result suggests that permutation polynomials over a ring of modulo $2^w$ are useful for cryptography.
\end{abstract}

\section{Introduction}
{
Vector Stream Cipher (VSC) is a stream cipher which was developed by one of the author at Communication Research Laboratory (now called the National Institute of Information and Communications Technology) in 2004 \cite{VSC}.
Although stream cipher is usually constructed by shift register and other parts based on finite field theory, VSC is constructed by permutation polynomials over a ring of modulo $2^w$ \cite{Rivest} and so is classified in chaos cipher.
For digital computers and digital signal processors, a modulo $2^w$ operation is practically negligible.
Therefore, the values of permutation polynomials over a ring of modulo $2^w$ can be calculated very fast, and so VSC is very fast.
The property of the polynomials is also useful for light implementation of VSC.

On the other hand, the security of VSC has been investigated by several researchers \cite{TL,A,B,Chosen}, and some security problems against for the theoretical attacks were reported though any practical attack breaking VSC has not been reported so far\cite{A,B,Chosen}.
In order to avoid the problems, the authors proposed in 2016 some improving of VSC and developed a new algorithm ``Vector Stream Cipher 2.0 (VSC 2.0)" \cite{Iwasaki-Umeno}.
Although VSC 2.0 is more secure than the original VSC, something to be optimized is left.
In this paper, we propose further improving of VSC as a result of optimization and  set it a new cipher called ``Vector Stream Cipher 2.1".

Permutation polynomials over a ring of modulo $2^w$ have not been applied for cryptography and not been studied except some applications and research \cite{Rivest,Knuth,Coveyou,RC6,One-stroke,Iwasaki-Umeno-arxiv}.
Our purpose of this research is not only to develop a new cipher but also to explore possibility of the permutation polynomials and chaos in the field of cryptography.

This paper is constructed as follows.
In sections 2 and 3, we introduce the algorithm of the original VSC and VSC 2.0, respectively.
In section 4, we propose improvements for VSC and VSC 2.0 and describe the optimized design as a new cipher ``Vector Stream Cipher 2.1".
In section 5, we perform some experiments to investigate the security of the methods.
Finally, we conclude this paper.
}

\section{Vector Stream Cipher}

In this section, we introduce Vector Stream Cipher 128 (VSC128) \cite{VSC} which is one kind of original VSC.
VSC128 requires 128-bit secret key and 128-bit initial vector.
The encryption algorithm is as follows: 

\begin{enumerate}
\item Assume that $A$, $B$, $C$, $D$, $X$, $Y$, $Z$ and $W$ are 32-bit integer variables.
Assign a secret key to $A$, $B$, $C$ and $D$, and an initial vector to $X$, $Y$, $Z$ and $W$.
\item Repeat the following operation 8 times. (In this paper, we call the operation ``round".)
\begin{enumerate}
\item Assume that $a$, $b$, $c$, $d$, $x$, $y$, $z$ and $w$ are 32-bit integer variables. 
Calculate the values of $a$, $b$, $c$, $d$, $x$, $y$, $z$ and $w$ as follows.
\begin{align*}
&a=A-(A \mod4)+1\mod2^{32},\\
&b=B-(B \mod4)+1\mod2^{32},\\
&c=C-(C\mod4)+1\mod2^{32},\\
&d=D-(D\mod4)+1\mod2^{32},\\
&x=X-(X\mod4)+1\mod2^{32},\\
&y=Y-(Y\mod4)+1\mod2^{32},\\
&z=Z-(Z\mod4)+1\mod2^{32},\\
&w=W-(W\mod4)+1\mod2^{32}.
\end{align*}
In this paper, we regard ``mod" as modulus operator. 
\item Assume {that} $A^\prime$, $B^\prime$, $C^\prime$, $D^\prime$, $X^\prime$, $Y^\prime$, $Z^\prime$ and $W^\prime$ {are} 32-bit integer variables. 

Calculate the values of $A^\prime$, $B^\prime$, $C^\prime$, $D^\prime$, $X^\prime$, $Y^\prime$, $Z^\prime$ and $W^\prime$ as follows.
\begin{align*}
&A^\prime=A(2A+y)\mod 2^{32},\\
&B^\prime=B(2B+x)\mod 2^{32},\\
&C^\prime=C(2C+z)\mod 2^{32},\\
&D^\prime=D(2D+w)\mod 2^{32},\\
&X^\prime=X(2X+c)\mod 2^{32},\\
&Y^\prime=Y(2Y+d)\mod 2^{32},\\
&Z^\prime=Z(2Z+a)\mod 2^{32},\\
&W^\prime=W(2W+b)\mod 2^{32}.
\end{align*}
\item Regard $(A^\prime,B^\prime,C^\prime,D^\prime,X^\prime,Y^\prime,Z^\prime,W^\prime)$ as a 256-bit sequence, and perform 5-bit left rotational shift.
After that, copy the sequence to $(A,B,C,D,X,Y,Z,W)$.
Writing mathematically,
\begin{align*}
&A=(A^\prime<<5)\oplus(B^\prime>>27)\mod2^{32},\ \ 
B=(B^\prime<<5)\oplus(C^\prime>>27)\mod2^{32},\\
&C=(C^\prime<<5)\oplus(D^\prime>>27)\mod2^{32},\ \ 
D=(D^\prime<<5)\oplus(X^\prime>>27)\mod2^{32},\\
&X=(X^\prime<<5)\oplus(Y^\prime>>27)\mod2^{32},\ \ 
Y=(Y^\prime<<5)\oplus(Z^\prime>>27)\mod2^{32},\\
&Z=(Z^\prime<<5)\oplus(W^\prime>>27)\mod2^{32},\ \ 
W=(W^\prime<<5)\oplus(A^\prime>>27)\mod2^{32}.
\end{align*}
Here, ``$<<$" means simple bit shift.
\end{enumerate}
\item Assume that $D1$, $D2$, $D3$ and $D4$ are
32-bit plaintexts and $E1$, $E2$, $E3$ and $E4$ are the corresponding ciphertexts respectively.
Then, calculate the values of $E1$, $E2$, $E3$ and $E4$ as follows.
\begin{align*}
&E1=D1\oplus X,\\
&E2=D2\oplus Y,\\
&E3=D3\oplus Z,\\
&E4=D4\oplus W.
\end{align*}
\item Repeat step 2 and 3 until all the given plaintexts are encrypted.
\end{enumerate}

\section{{VSC 2.0}}

There are some security problems of VSC128.
We think that the following {three} points are the most important problems.
\begin{itemize}
\item The maximum linear characteristic probability of VSC128 is $2^{-115}$.
Therefore, {distinguishing attack with linear masking} is practical \cite{A,B}.
\item It is reported that the output sequence (key-stream) of VSC128 have a statistical deviation if the initial vector is chosen among specific vectors, and so the distinguishing attack is realized if an attacker can chose an initial vector intentionally \cite{Chosen}.
\item The round of VSC128 is not a bijection.
Then, the effective key length of VSC128 is smaller than 128bit.
\end{itemize}

To solve the above problems, we proposed ``Vector Stream Cipher 2.0 (VSC 2.0)" \cite{Iwasaki-Umeno}, which has the following three improvements.
\begin{itemize}
\item Changing the iteration number of the round from 8 to 9. 
Herewith, the maximum linear characteristicprobability changes from $2^{-115}$ to $2^{-129}$, and so a distinguishing attack with linear masking becomes not practical.
\item Adding a preprocessing which replaces a given initial vector with another value like a hash value.
Herewith, an attacher cannot choose an initial vector intentionally.
\item Introducing a new rule ``Keep the value of the variable $D$ is even".
Herewith, by the following Theorem \ref{2.0thm}, the round becomes a bijection.
In order to keep the new rule, the key length and step 2(c) of the VSC128 algorithm change.
\end{itemize}

\begin{thm}
\label{2.0thm}
Consider a map $g:(\mathbb{Z}/2^n\mathbb{Z})^m\to(\mathbb{Z}/2^n\mathbb{Z})^m$, which is described as
\begin{align*}
&g(A_0,A_1,\cdots,A_{m-1})=(A_0^\prime,A_1^\prime,\cdots,A_{m-1}^\prime),\\
&A_i^\prime=A_i\left(2A_i+a\left(A_{\left(i+1\mod m\right)}\right)\right) \mod 2^n\ \ (^\forall i\in\mathbb{Z}/2^m\mathbb{Z}),
\end{align*}
where $A_1,\cdots,A_m$ and $A_1^\prime,\cdots,A_m^\prime$ are elements of $\mathbb{Z}/2^n\mathbb{Z}$ and $a(A_i)=A_i-(A_i\mod4)+1$ $(^\forall i \in \mathbb{Z}/2^m\mathbb{Z})$.
Assume that $O_{n}$ is a subset of $\mathbb{Z}/2^n\mathbb{Z}$, which is constructed of odd numbers in $\mathbb{Z}/2^n\mathbb{Z}$.
Then, if we restrict the domain of $g$ to $(\mathbb{Z}/2^n\mathbb{Z})^m$ except $(O_n)^m$, $g$ become a bijective map on $(\mathbb{Z}/2^n\mathbb{Z})^m\backslash(O_n)^m$.
\end{thm}
\noindent If you know the proof of Theorem {\ref{2.0thm}}, see Ref. \cite{Iwasaki-Umeno}.

The algorithm of VSC 2.0 which is based on the above is as follows:
\begin{enumerate}
\item Assume that $A$, $B$, $C$, $D$, $X$, $Y$, $Z$ and $W$ are 32-bit integer variables.
Set $A$=0xfedcba98, $B$=0x01234567, $C$=0x89abcdef and $D$=0x76543210 and assign a secret key to $A$, $B$, $C$ and $D$, and an initial vector to $X$, $Y$, $Z$ and $W$.
\item Repeat the following operation {30}  times. (The operation is the ``round" of VSC 2.0.)
\begin{enumerate}
\item Assume that $a$, $b$, $c$, $d$, $x$, $y$, $z$ and $w$ are 32-bit integer variables. 
Calculate the values of $a$, $b$, $c$, $d$, $x$, $y$, $z$ and $w$ as follows.
\begin{align*}
&a=A-(A \mod4)+1 \mod 2^{32},\\
&b=B-(B \mod4)+1 \mod 2^{32},\\
&c=C-(C\mod4)+1 \mod 2^{32},\\
&d=D-(D\mod4)+1 \mod 2^{32},\\
&x=X-(X\mod4)+1 \mod 2^{32},\\
&y=Y-(Y\mod4)+1 \mod 2^{32},\\
&z=Z-(Z\mod4)+1\mod 2^{32},\\
&w=W-(W\mod4)+1 \mod 2^{32}.
\end{align*}
\item Assume {that} $A^\prime$, $B^\prime$, $C^\prime$, $D^\prime$, $X^\prime$, $Y^\prime$, $Z^\prime$ and $W^\prime$ {are} 32-bit integer variables. 

Calculate the values of $A^\prime$, $B^\prime$, $C^\prime$, $D^\prime$, $X^\prime$, $Y^\prime$, $Z^\prime$ and $W^\prime$ as follows.
\begin{align*}
&A^\prime=A(2A+y)\mod 2^{32},\\
&B^\prime=B(2B+x)\mod 2^{32},\\
&C^\prime=C(2C+z)\mod 2^{32},\\
&D^\prime=D(2D+w)\mod 2^{32},\\
&X^\prime=X(2X+c)\mod 2^{32},\\
&Y^\prime=Y(2Y+d)\mod 2^{32},\\
&Z^\prime=Z(2Z+a)\mod 2^{32},\\
&W^\prime=W(2W+b)\mod 2^{32}.
\end{align*}
\item Regard $(A^\prime,B^\prime,C^\prime,D^\prime,X^\prime,Y^\prime,Z^\prime,W^\prime)$ as a 256-bitssequence, and perform 5-bit left rotational shift.
After that, copy the sequence to $(A,B,C,D,X,Y,Z,W)$.
After that, 1-bit left rotational shift for low-ranking 6-bit of $D$. 
Writing mathematically,
\begin{align*}
&A=(A^\prime<<5)\oplus(B^\prime>>27) \mod 2^{32},\ \ 
B=(B^\prime<<5)\oplus(C^\prime>>27)\mod 2^{32},\\
&C=(C^\prime<<5)\oplus(D^\prime>>27)\mod 2^{32},\ \ 
D=(D^\prime<<5)\oplus((X^\prime>>27)<<1))\mod 2^{32},\\
&X=(X^\prime<<5)\oplus(Y^\prime>>27) \mod 2^{32},\ \ 
Y=(Y^\prime<<5)\oplus(Z^\prime>>27) \mod 2^{32},\\
&Z=(Z^\prime<<5)\oplus(W^\prime>>27) \mod 2^{32},\ \ 
W=(W^\prime<<5)\oplus(A^\prime>>27) \mod 2^{32}.
\end{align*}
\end{enumerate}
\item Assign a secret key to $A$, $B$, $C$ and $D$ except the least significant bit of $D$.
Set the least significant bit of $D$ to 0.
\item Perform the round 9 times.
\item Assume {that} $D1$, $D2$, $D3$ and $D4$ {are} 32-bit plaintexts and $E1$, $E2$, $E3$ and $E4$ are the corresponding ciphertexts respectively.
Then, calculate the values of $E1$, $E2$, $E3$ and $E4$ as follows.
\begin{align*}
&E1=D1\oplus X,\\
&E2=D2\oplus Y,\\
&E3=D3\oplus Z,\\
&E4=D4\oplus W.
\end{align*}
\item Repeat step 4 and 5 until all the given plaintexts are encrypted.
\end{enumerate}

\section{Further improvements of VSC 2.0}

Although the security of VSC 2.0 is better than the original VSC, something to be considered for optimization of cipher design is left.
In particular, we think that the following two things should  be considered for further improvements.
\begin{itemize}
\item The key length of VSC 2.0 is 127bit.
Thus, the key space is the half of the original VSC128.
\item The step 2(c) of VSC 2.0 algorithm is slower than that of VSC128.
\end{itemize}
Both of them are caused by what makes the round a bijection.
To improve {them}, we introduce the following theorem.

\begin{thm}
\label{2.1thm}
Assume that $g:(\mathbb{Z}/2^n\mathbb{Z})^m\to(\mathbb{Z}/2^n\mathbb{Z})^m$ is described as
\begin{align*}
&g(A_0,A_1,\cdots,A_{m-1})=(A_0^\prime,A_1^\prime,\cdots,A_{m-1}^\prime),\\
&A_i^\prime=A_i\left(2A_i+4A_{\left(i+1\mod m\right)}+1\right) \mod 2^n\ \ (^\forall i\in\mathbb{Z}/2^m\mathbb{Z}).
\end{align*}
Here, $A_1,\cdots,A_m$ and $A_1^\prime,\cdots,A_m^\prime$ are elements of $\mathbb{Z}/2^n\mathbb{Z}$.
Then, $g$ is a bijection.
\end{thm}

\begin{pro}
It is enough to prove injectivity.
Assume that $(A_0, A_1,\cdots,A_{m-1})$ and $(\tilde{A_0}, \tilde{A_1},\cdots,\tilde{A_{m-1}})$ are elements of $(\mathbb{Z}/2^n\mathbb{Z})^m$ satisfying
\[(A_0, A_1,\cdots,A_{m-1})\ne(\tilde{A_0}, \tilde{A_1},\cdots,\tilde{A_{m-1}}).\]
There are non-negative numbers $s_i$ $(i\in\mathbb{Z}/2^m\mathbb{Z})$ satisfying
\begin{align*}
& s_i\leq n\ \  ( ^\forall i\in\mathbb{Z}/2^m\mathbb{Z}),\\
&(s_0,s_1,\cdots,s_{m-1})\ne(n,n,\cdots,n),\\
&\tilde{A_i}=A_i+(2p_i-1)2^{s_i}\mod2^n\ \  ( ^\forall i\in\mathbb{Z}/2^m\mathbb{Z}),
\end{align*}
where $p_i$ are natural numbers.
Assume that
\begin{align*}
(A_0^\prime,A_1^\prime,\cdots,A_{m-1}^\prime)=&g(A_0,A_1,\cdots,A_{,-1}),\\
(\tilde{A_0}^\prime,\tilde{A_1}^\prime,\cdots,\tilde{A}_{m-1}^\prime)=&g(\tilde{A_0},\tilde{A_1},\cdots,\tilde{A}_{m-1}).
\end{align*}
Then, there exists $k\in\mathbb{Z}/2^m\mathbb{Z}$ such that $s_k\leq s_i$ $( ^\forall i\in\mathbb{Z}/2^m\mathbb{Z})$.
Obviously, $s_k<n$.
To simplify notation,  we define new symbols $A_m$, $\tilde{A_m}$, $s_m$ and $p_m$ as $A_0$, $\tilde{A_0}$, $s_0$ and $p_0$, respectively.
Then, 
\begin{align*}
&\ \ \ \ \tilde{A_k}^\prime\\&=\tilde{A_k}\{2\tilde{A_k}+4\tilde{A}_{k+1}+1\} \mod2^n\nonumber\\
&=\{A_k+(2p_k-1)2^{s_k}\}\{2A_k+(2p_k-1)\cdot2^{s_k+1}+4A_{k+1}+(2p_{k+1}-1)\cdot2^{s_{k+1}+2}+1\} \mod 2^n\nonumber\\
&=A_k^\prime+(2p_k-1)\cdot2^{s_k}+r\cdot2^{s_k+1}+r^\prime\cdot2^{s_{k+1}+2}\mod 2^n.
\end{align*}
Here, $r$ and $r^\prime$ are natural numbers.
Since $s_k\leq s_{k+1}$ and $s_k<n$, $\tilde{A_k}^\prime\ne A_k^\prime$.
From the above, $g$ is a bijection.\qed
\end{pro}

By using Theorem \ref{2.1thm}, we can further improve VSC 2.0.
We propose a new cipher ``Vector Stream Cipher 2.1 (VSC 2.1)".
The algorithm proposed here as VSC 2.1 is as follows:

\begin{enumerate}
\item Assume that $A$, $B$, $C$, $D$, $X$, $Y$, $Z$ and $W$ are 32-bit integer variables.
Set $A$=0xfedcba98, $B$=0x01234567, $C$=0x89abcdef and $D$=0x76543210 and assign a secret key to $A$, $B$, $C$ and $D$, and an initial vector to $X$, $Y$, $Z$ and $W$.
\item Repeat the following operation 30 times. (The operation is the ``round" of VSC 2.1.)
\begin{enumerate}
\item Assume that $a$, $b$, $c$, $d$, $x$, $y$, $z$ and $w$ are 32-bit integer variables. 
Calculate the values of $a$, $b$, $c$, $d$, $x$, $y$, $z$ and $w$ as follows.
\begin{align*}
&a=4A+1\mod 2^{32},\\
&b=4B+1\mod 2^{32},\\
&c=4C+1\mod 2^{32},\\
&d=4D+1\mod 2^{32},\\
&x=4X+1\mod 2^{32},\\
&y=4Y+1\mod 2^{32},\\
&z=4Z+1\mod 2^{32},\\
&w=4W+1\mod 2^{32}.
\end{align*}
\item Assume {that} $A^\prime$, $B^\prime$, $C^\prime$, $D^\prime$, $X^\prime$, $Y^\prime$, $Z^\prime$ and $W^\prime$ {are} 32-bit integer variables. 

Calculate the values of $A^\prime$, $B^\prime$, $C^\prime$, $D^\prime$, $X^\prime$, $Y^\prime$, $Z^\prime$ and $W^\prime$ as follows.
\begin{align*}
&A^\prime=A(2A+y)\mod 2^{32},\\
&B^\prime=B(2B+x)\mod 2^{32},\\
&C^\prime=C(2C+z)\mod 2^{32},\\
&D^\prime=D(2D+w)\mod 2^{32},\\
&X^\prime=X(2X+c)\mod 2^{32},\\
&Y^\prime=Y(2Y+d)\mod 2^{32},\\
&Z^\prime=Z(2Z+a)\mod 2^{32},\\
&W^\prime=W(2W+b)\mod 2^{32}.
\end{align*}
\item Regard $(A^\prime,B^\prime,C^\prime,D^\prime,X^\prime,Y^\prime,Z^\prime,W^\prime)$ as a 256-bit sequence, and perform 5-bit left rotational shift.
After that, copy the sequence to $(A,B,C,D,X,Y,Z,W)$.
Writing mathematically,
\begin{align*}
&A=(A^\prime<<5)\oplus(B^\prime>>27) \mod 2^{32},\ \ 
B=(B^\prime<<5)\oplus(C^\prime>>27) \mod 2^{32},\\
&C=(C^\prime<<5)\oplus(D^\prime>>27) \mod 2^{32},\ \ 
D=(D^\prime<<5)\oplus(X^\prime>>27)) \mod 2^{32},\\
&X=(X^\prime<<5)\oplus(Y^\prime>>27) \mod 2^{32},\ \ 
Y=(Y^\prime<<5)\oplus(Z^\prime>>27) \mod 2^{32},\\
&Z=(Z^\prime<<5)\oplus(W^\prime>>27) \mod 2^{32},\ \ 
W=(W^\prime<<5)\oplus(A^\prime>>27) \mod 2^{32}.
\end{align*}
\end{enumerate}
\item Assign a secret key to $A$, $B$, $C$ and $D$.
\item Perform the round 9 times.
\item Assume {that} $D1$, $D2$, $D3$ and $D4$ are 32-bit plaintexts and $E1$, $E2$, $E3$ and $E4$ are the corresponding ciphertexts respectively.
Then, calculate the values of $E1$, $E2$, $E3$ and $E4$ as follows.
\begin{align*}
&E1=D1\oplus X,\\
&E2=D2\oplus Y,\\
&E3=D3\oplus Z,\\
&E4=D4\oplus W.
\end{align*}
\item Repeat step 4 and 5 until all the given plaintexts are encrypted.
\end{enumerate}
By Theorem \ref{2.1thm}, the round of VSC 2.1 is a bijection.
The maximum linear probability of VSC 2.1 is the same as that of VSC 2.0.
The key length of VSC 2.1 is 128bit, it is longer than that of VSC 2.1.
VSC 2.1 is expected to be faster than VSC 2.0 because step 2(c) is more simple than that of VSC 2.0.

\section{Experiments} 

In this section, we perform some experiments for VSC 2.1.

\subsection{Speed}

We measure the speeds of performing VSC128, VSC 2.0, VSC 2.1 and AES-128.
The environment in which we measure is shown in Table \ref{A}.
As results, we got Table \ref{B}.
VSC2.1 is slightly slower than the original VSC128, but faster than VSC 2.0.
\begin{table}[h]
\begin{center}
\caption{Environment on which we measure speed}
\label{A}
\begin{tabular}[t]{|c|c|}
\hline
CPU& 1.3 GHz Intel Core i5\\
\hline
Memory & 4 GB 1600 MHz DDR3\\
\hline
OS & OS X 10.9.5（13F34）\\
\hline
Compiler& gcc 4.2.1\\
\hline
Optimization option& -Ofast\\
\hline
\end{tabular}
\end{center}
\end{table}
\begin{table}[h]
\begin{center}
\caption{Encryption speeds}
\label{B}
\begin{tabular}[t]{|c|c|}
\hline
Algorithm& Speed(Mbps)\\
\hline
\hline
VSC128 & 1202.254889\\
\hline
VSC 2.0& 1039.222464\\
\hline
VSC 2.1& 1113.193866\\
\hline
AES-128 ECB &366.901621\\
\hline
\end{tabular}
\end{center}
\end{table}

\subsection{Property of the preprocessing}

We investigate properties of the preprocessing of VSC 2.0 and VSC 2.1.
Step 2 of VSC 2.0 algorithm and that of VSC 2.1 algorithm are preprocessing, respectively.
The detail of the experiment is as follows:

 \begin{enumerate}
\item Select an input randomly. (We call the input $I_1$.)
\item Select a bit of $I_1$ and reverse the bit. (We call the value $I_2$.)
\item Apply the preprocessing to $I_1$ and $I_2$. (We call the outputs $I_1^\prime$ and $I_2^\prime$ respectively.)
\item Measure the Haming distance between $I_1^\prime$ and $I_2^\prime$.
\item Repeat step 1-4 1000000 times. Calculate the average of the Haming distance which are measure at step 3.
\end{enumerate}
As a result, we got Table \ref{ham}.
Since the output length of the both preprocessing are 128bit, the result shows that the preprocessings have a good property.
\begin{table}[h]
\begin{center}
\caption{} \label{ham}
\begin{tabular}[t]{|c|c|}
\hline
Algorithm& Average Haming distance\\
\hline
\hline
VSC 2.0& 64.000107\\
\hline
VSC 2.1& 63.995965\\
\hline
\end{tabular}
\end{center}
\end{table}

\subsection{Randomness of key stream}

We performed randomness test described by NIST SP800-22 \cite{NIST} for key streams generated by VSC128, VSC 2.0 and VSC 2.1.
The test was performed for 11 sets.
Each set is constructed of 1000 sequences. (Exceptionally, the sets 10 and 11 are constructed of 255 sequences respectively. A sequence of the set 10 is generated with an initial condition (key and initial vector) whose one bit is ``1" and the others are ``0". VSC 2.0 requires the least bit of $D$ is ``0". 
Then, the set 10 is constructed of only 255 sequences.
A sequence of the set 11 is generated with an initial condition whose two bits are ``0" and the others are ``1".
One of the two is the least bit of $D$.
Then, set 11 is also constructed of only 255 sequences.
)
Each sequence is constructed of 1000000bits, which are generated by {VSC128, VSC 2.0 or VSC 2.1} with a secret key and an initial vector.
 The secret key and the initial vector are chosen randomly, but random pattern is dependent on a set.
Table \ref{C} shows the result.
The randomness test is constructed of 188 test items.
Even if sequences are exactly random, there are cases that the sequence does not pass all test items.
Therefore, the result show that there are no problem about randomness of the key stream of VSC128, VSC 2.0 and VSC 2.1. 

\begin{table}[htbp]
\begin{center}
\caption{Results of randomness tests}
\label{C}
\begin{tabular}[t]{|c|c|c|c|}
\hline
&\multicolumn{3}{|c|}{Numbers  of test items which the set passed}\\
\cline{2-4}
Set No. &\ \  \ VSC128\ \  \ &\ \ \  VSC 2.0\ \ \  &VSC 2.1\\
\hline
\hline
1&188&188&187\\
\hline
2&188 &187&188\\
\hline
3&188 &186&188\\
\hline
4&187 &188&188\\
\hline
5&187 &188&188\\
\hline
6&188 &187&188\\
\hline
7&188 &187&188\\
\hline
8&188 &188&187\\
\hline
9&188 &188&188\\
\hline
10&188 &188&188\\
\hline
11&188 &188&188\\
\hline

\end{tabular}
\end{center}
\end{table}

\section{Conclusion}

We proposed further improving of VSC as a certain optimization of cipher design.
Our proposed VSC 2.1 is faster than VSC 2.0.
We think that VSC 2.1 is more secure than VSC 2.0 because of the following two reasons.
First is that the key length of VSC 2.1 is more long than that of VSC 2.0.
Second is that any theoretical attacks which were reported as workable attacks for the original VSC128 are not workable for VSC 2.1.
In particular, VSC 2.1 has the provable security for the distinguishing attack with linear masking.
Thus, VSC 2.1 is a precious example of chaos cipher which is using permutation polynomials over a ring of modulo $2^w$, and suggests that the permutation polynomials and chaos theory are useful for cryptography.

\end{document}